\documentclass[10pt]{RoE} 

\usepackage{graphicx}
\usepackage{float}
\usepackage{hyperref}
\usepackage{listings}
\usepackage[utf8]{inputenc}
\usepackage[english]{babel}

\graphicspath{{images/}} 

\JournalInfo{TBD} 

\PaperType{Vision paper} 
\DOI{DOI: xxxxxxxxxxxxx} 

\PaperTitle{The Future of Intelligent Wavefront Shaping for Smart Radio Environments} 

\Authors{Benjamin W. Frazier\textsuperscript{1,2,3}* and Steven M. Anlage\textsuperscript{2,4,5}} 
\affiliation{\textsuperscript{1}\textit{Institute for Research in Electronics and Applied Physics, University of Maryland, College Park, MD 20742, USA}} 
\affiliation{\textsuperscript{2}\textit{Department of Electrical and Computer Engineering, University of Maryland, College Park, MD 20742, USA}} 
\affiliation{\textsuperscript{3}\textit{Johns Hopkins University Applied Physics Laboratory, Laurel, MD 20723, USA}} 
\affiliation{\textsuperscript{4}\textit{Department of Physics, University of Maryland, College Park, MD 20742, USA}} 
\affiliation{\textsuperscript{5}\textit{Quantum Materials Center, University of Maryland, College Park, MD 20742, USA}} 
\affiliation{*\textbf{Corresponding author}: Benjamin.Frazier@jhuapl.edu} 


\newcommand{\keywordname}{Key terms} 

\Dates{Submitted Jan. 26, 2020}

\Abstract{As the electromagnetic spectrum becomes more congested and the environments in which we need to operate become more complicated, control over the environment itself becomes necessary to ensure the integrity of wireless communication channels. Wavefront shaping with programmable metasurfaces allows wave fields to be manipulated in both time and space, providing a method to interact with the environment. When coupled with deep learning, intelligent wavefront shaping serves as a catalyst, enabling smart radio environments and unlocking applications beyond traditional wireless communication networks. In this paper, we discuss the outlook of intelligent wavefront shaping for wave propagation in complex environments and highlight its transformative potential.}

\begin{document}
\flushbottom 
\maketitle
Modern radio frequency (RF) imaging and communications systems operate in complex environments that are susceptible to multipath reflections which scramble propagating electromagnetic waves. Transmitted signals in these environments experience random spatio-temporal fluctuations which degrade or disrupt performance, particularly when combined with competing RF emissions and a congested electromagnetic spectrum. A ``smart'' radio environment (SRE) must be able to handle such conditions, adapting on-the-fly to optimize a metric for the wireless channel \cite{Reference1,Reference2}. The optimization should be performed over the entire propagation path, not only at the endpoints as with traditional wireless systems. 

\begin{figure}[ht]\centering
\includegraphics[width = .95\linewidth]{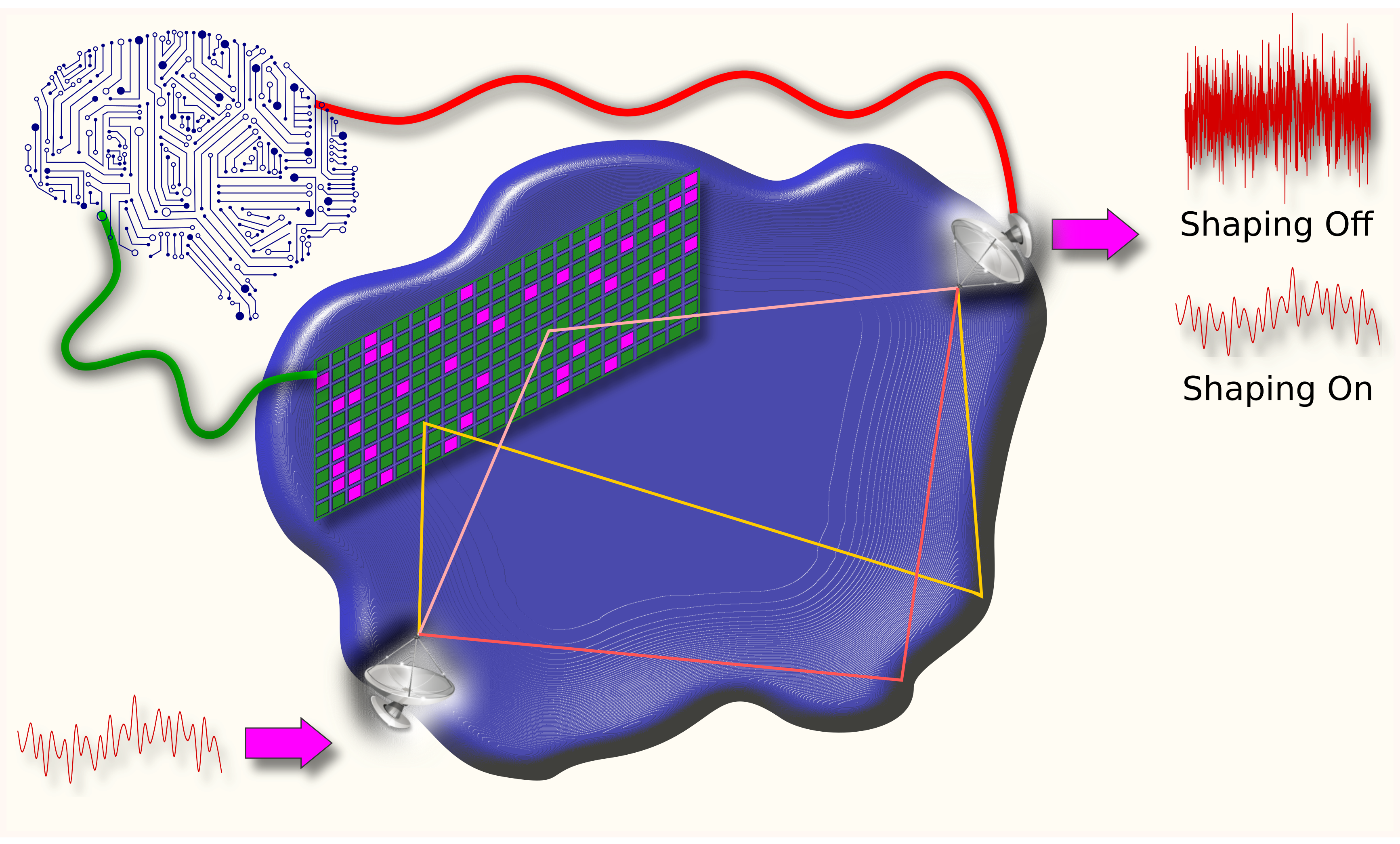}
\caption{\label{fig:overview} \bf Conceptual diagram of a smart radio environment (SRE). \normalfont A modulated signal input to a complex environment, such as a wireless network, is scrambled through constructive and destructive interference between the multiple paths. A reconfigurable metasurface is then leveraged to program the environment through intelligent wavefront shaping, reducing the interference and allowing the input signal to be recovered. Deep learning provides the intelligence, updating the metasurface and responding to changes in the environment on-the-fly.}
\end{figure}

The vision of an SRE (Fig. \ref{fig:overview}) is a self-adaptive system that can counter scrambling of electromagnetic waves from the complex scattering environment, ensuring operation even under degraded conditions. The ability to program the environment can be achieved through the use of tunable metasurfaces, which can manipulate their local surface impedance to enable on-demand beamforming \cite{Reference3}; these devices are inexpensive and widely available at RF wavelengths, making them ideal for dynamic wavefront shaping applications.

It is natural to consider the wireless channel for an SRE as a long-range, open-world environment that may be a densely packed urban area or a sparsely populated farm. However, an SRE is also valuable in enclosed environments, such as a train station, a passenger compartment on an airplane, or even an office. An SRE therefore has many potential applications, including the ability to enhance 5G communications, protect against electromagnetic interference, induce cold spots at specified locations \cite{Reference4}, realize a microwave cloak around an object \cite{Reference5}, enable computational imaging \cite{Reference6}, and leverage Wi-Fi signals to allow wireless backscatter communications \cite{Reference7}. 

Complex microwave cavities can mimic these larger scale enclosures and are extremely useful as surrogates for prototyping and experimentation \cite{Reference8}. Many metasurface enabled wavefront shaping experiments have been performed in these cavities \cite{Reference9,Reference10,Reference11,Reference12}, demonstrating fine control over the scattering parameters. As such, complex microwave cavities will play a crucial role in the future development of SREs, with intelligent wavefront shaping serving as an enabling technology.

The metasurface is placed inside the scattering medium, intercepting a relatively small number of ray trajectories, sometimes with multiple bounces off the metasurface itself. Therefore, the relationship between metasurface commands and sensed environmental responses is extremely complex, resulting in an ill-posed inverse problem; an issue that is exacerbated when considering multiple distributed surfaces. Central to the vision is intelligence, meaning the device must sense the environment and then deliberately interact with it. The explosion of software defined radios in recent years provides a wealth of powerful and inexpensive hardware to develop sensing prototypes. The distinction of the architecture as ``software defined'' means that general purpose hardware can be reused on many varied applications, significantly reducing cost. 

Wavefront shaping applications to date have relied on brute force trial and error or stochastic search algorithms \cite{Reference11,Reference13}; however, the inherent complexity makes this an ideal place to leverage deep learning. Deep learning has enjoyed great success in the design of metasurfaces, but has had less use in dynamic wavefront shaping applications. To successfully field a deep learning solution, we need to address concerns with long processing times as well as size, weight, and power. Traditional deep learning systems require tremendous amounts of data to train. The acquisition time for sufficient training data may be longer than the coherence time of the environment \cite{Reference2}, resulting in a trained solution that is no longer accurate. The concept of reinforcement learning \cite{Reference14} is at the intersection of deep learning and optimal control, and can assist here. It provides a methodology for adjusting the SRE to environmental changes on-the-fly, and has shown great potential to develop optimal control policies in complex and uncertain environments. To alleviate concerns with overwhelming amounts of data, reinforcement learning can be coupled with transfer learning, where information about previous environments is leveraged to accelerate training \cite{Reference15}.

Processing for deep learning is typically performed on expensive and power hungry graphical processing units, so the footprint in terms of both cost and power may exceed allowable margins. Computational efficiency can be increased by compressing deep learning models through quantization and pruning \cite{Reference16}. The growth of edge intelligence for connected devices in the internet of things (IoT) has produced a demand for smaller deep learning models and cheaper, more efficient processing, which will lead to a wider availability of viable processing platforms.

Cabling and interface requirements grow with the number of unit cells provided by a metasurface. The ability to address individual unit cells and switch states as needed is critical to achieve a practical SRE. Connectors and large cable runs form bottlenecks and tend to be the weakest links in a system, so the capability of addressing unit cells wirelessly without further corrupting the environment is highly desired for large element counts.

Future wireless systems, including 5G and IoT devices, will increasingly rely on the ability to program the environment as the electromagnetic spectrum becomes even more congested. Intelligent wavefront shaping using metasurfaces coupled with deep learning serves as a path towards realizing an SRE, which will be a revolutionary breakthrough for imaging and communication systems operating in complex environments.


\end{document}